\documentclass[%
reprint,
amsmath,amssymb,
aps,
]{revtex4-2}
\usepackage{graphicx}
\usepackage{dcolumn}
\usepackage{bm}
\usepackage{xcolor}

\begin{document}
	

	\title{Design and Stability Analysis of an Electromechanical Model for Nerves}
	
	\author{F. M. Moukam Kakmeni}%
	\altaffiliation{Corresponding Author, Email:
		moukamkakmeni@gmail.com}
	\affiliation{Complex Systems and Theoretical Biology Group
		(CoSTBiG),\\ Laboratory of Research on Advanced Materials and
		Nonlinear Science~(LaRAMaNS), Department
		of Physics, Faculty of Science, University of Buea, PO Box 63, Buea, Cameroon}%

	\author{B. Njinabo Akoni}
	\affiliation{Complex Systems and Theoretical Biology Group (CoSTBiG),\\ Laboratory of Research on Advanced Materials and Nonlinear Science~(LaRAMaNS), Department
		of Physics, Faculty of Science, University of Buea, PO Box 63, Buea, Cameroon}

	\date{\today}
%
\begin{abstract}

In this work, we propose and study the stability of nerve impulse propagation as  electrical and mechanical signals through linear approximation. We present a potential energy stored in the biomembrane due to the deformation, bending, and stretching as the action potential propagates in the nerve fibre. From the potential energy, we derive electromechanical coupling forces and an attempt is made to unify the two models to account for both the electrical and mechanical activities of nerve signal propagation by introducing the electromechanical coupling forces. We examine the stability of the equilibrium states of the electromechanical model for nerves through the Routh Hurwitz stability criteria. Finally, we present results of the numerical simulations of the electromechanical model for nerves through Runge Kutta method of order four.

\end{abstract}
%

%
%
%
\maketitle

\section{Introduction}
\label{sec1} 

For a long time, researchers have theoretically and experimentally investigated how nerve pulse propagation works  \cite{ s1, s11,s12, s2, s3,  F1, F2,  F3, F4, F10, F11, F12}. In the early 1950 s, Hodgkin and Huxley presented a model with which they provided a quantitative description of the electrical events underlying the generation and propagation of a nerve impulse  \cite{s1}. The Hodgkin-Huxley (HH) model was the centered study electrophysiology that had started in 1791 with the work of Galvani  \cite{s11} and later by Bernstein's pioneer idea  \cite{s12} of a key role held by membrane's permeability in ions flow across the nerve cell membrane. In the standard model, introduced by Hodgkin and Huxley (HH) \cite{s1}, the axon, which is a part of the nerve is basically treated as an electrical cable.  The electrical signal is generated and propagated along the nerve due to  nonlinear effects as a result of the opening and closing of ion channels proteins  located in the cell membrane. It creates a feedback mechanism  allowing the signal to propagate over macroscopic distances in the nerve fibre .

The above description of the action potential based on the Hodgkin-Huxley (HH) model is limited because it mainly assumes that the membrane capacitance is fixed while the ionic current results from protein channels which is a complete different assumption made in the soliton model \cite{h3, hh1, h1, h2}. However, experimental results suggested that the changes in membrane capacitance are of an order of magnitude sufficient to account for the observed voltage changes during the action potential (AP) \cite{h3} . This variation in membrane capacitance may emanate from changes in thickness and area (among other physiological factors) of the membrane during the action potential  \cite{hh1}. Furthermore, it is important to note that in the complete absence of proteins, lipid membrane form  voltage-gated pores  display properties indistinguishable from the protein channels \cite{h1, h2}. These ions channels are said to originate from area fluctuations in the lipid membranes. At melting transitions,  membrane permeability is reported to be  maximum with membrane very susceptible to piezoelectric effects ( this is a phenomenon whereby an electric charge accumulates in certain solid materials in response to applied mechanical stress.) \cite{h2}.  This is important because biological membranes under physiological conditions in fact exist in a state close to melting transitions  , and lipid based channels can be expected under physiological conditions \cite{F4, F10, h2}.

The behaviour of the HH model for nerve is in good agreement with several electrical properties of the propagated nerve impulse in experiments. However, the model cannot account for non-electrical manifestations of the nerve impulse for which there is ample experimental evidence. Heimburg and Jackson \cite{F4, F10}  have proposed that the AP is a propagating density pulse (soliton), and therefore an electromechanical rather than a purely electrical phenomenon. The soliton model is based on the thermodynamics and phase behaviour of the lipids in the cell membrane. A soliton or solitary pulse is known to be a localized pulse propagating without attenuation and without change of shape. In mathematical physics, two conditions are necessary for the existence of solitons: the pulse speed should be frequency dependent and a non-linear function of the pulse amplitude. Within lipid phase transition, both conditions for existence of solitons: dispersion and non-linearity of the speed of sound are present and a soliton can propagate along the axon membrane. In soliton theory, the electrical component of nerve signals originates from piezoelectric effects and lipid ions voltage-gated conductions at phase melting transitions. Piezoelectricity and flexoelectricity are thought to result from both the large electric dipole moments of the lipid molecules and from the asymmetry in the distribution of negatively charged lipids (typically about 10 \%) found primarily in the inner leaflet of the bilayer \cite{hh1, h2}.

Currently, there is  an active debate whether the soliton  model which is supported by experimental measurements in artificial lipid membranes  can replace the HH model  \cite{p1}.  In contrary, instead of contrasting these two models of  nerve impulse propagation, Engelbrecht \emph{et al.}  \cite{p2}  presented a model that seek to unify them. Their model  is developed to incorporate, integrate and explain all relevant aspects of the nerve impulse by unifying different manifestations of the nerve impulse and the interactions between them. Such a complex system is characterized by the interaction of the electrical and mechanical components of the nerve signals. The interaction of these signals brings about a new quality at the macrolevel. The Engelbrecht model, is developed to unify all relevant manifestations of the nerve pulse propagation and  their interaction(s) but it is worthy to point out that the mechanism of coupling is not satisfactory due to the distinct purposes for which they are developed and conflicting assumptions these purposes often require.

In the present work, we seek to design an electromechanical model unifying the relevant manifestations of the nerve pulse propagation by introducing external coupling forces and analysing their effect on the stability of the electromechanical model. This idea is base on the fact that biological systems are characterised by many interwoven properties and interactions between the various phenomenon play a decisive role. While measuring and understanding the electrical component of the AP has been focus on most experimental and theoretical efforts, a large number of experimental studies have shown that the AP is accompanied by mechanical changes. Our motivation is base on the fact that the electrically driven mechanical modes that we consider are surface waves in which potential energy is stored in the elastic energy associated with deforming biomembrane.

The next section (II) of this work is to design a model that will be able to describe waves of two physical origins(electrical and mechanical) in which the nonlinearity plays a decisive role. The simplest idea is to base the 'coherent theory' on existing mathematical models of FitzHugh Nagumo  describing the electrical activities of the nerve(action potential) and the mechanical wave in the surrounding biomembrane governed by the improved Heimburg Jackson model equation \cite{f1}  but introducing the coupling forces \cite{p2, b34, b35}. In section III, we shall study the stability of the equilibrium states of the electromechanical model of nerves through the Routh-Hurwitz stability criterion. Section IV is devoted to numerical result in order to support our analysis. Finally, we shall present a summary of the work in the last section.

\section{The Model}
 Biomembranes are able to resist pressure, tension, stretch, and bending \cite{b31}. The deformation can be induced by an electrical or  a mechanical impact \cite{f2, b32}. For dynamical equations of the electrical singnal, we make use of the general form of the classical two dimensional FitzHugh and Nagumo system. The  mechanical responses been predicted and well accounted for by the so called Heimburg model of the nerve. In this last model, the nerve impulse is regarded as a thermodynamic phenomenon related to the characteristic properties of the membrane \cite{b33}.

Thus, let us consider the action potential is governed by the following equations:
\begin{eqnarray}\label{m110} 
\frac{\partial V}{\partial t}&=&\frac{1}{C}\left [\frac{1}{\rho}\left( V-\frac{1}{3}\frac{V^{3}}{V_{0}^{2}}\right) +i_{L}\right]+D_{1}\frac{\partial^{2} V}{\partial x^{2}}\nonumber\\  
\dfrac{\partial i_{L}}{\partial t}&=&\frac{1}{L}(-V-Ri_{L}+E)
\end{eqnarray}
for the electrical part, where $V$ is the voltage across the membrane at time $t$, $V_{0}$ is the initial voltage across the membrane, $D_{1}$ is  the  dispersion coefficient, $i_{l}$ is the recovery current, and $C$ is the membrane capacitance. For the mechanical part by using the improved Heimburg and Jackson model with dissipative term \cite{f1}, the longitudinal wave  in the biomembrane is governed by the equations:
\begin{eqnarray}\label{m111} 
\frac{\partial^{2}}{\partial t^{2}}\Delta\rho^{A}&=&\frac{\partial }{\partial x}\left [(c^{2}_{0}+\alpha\Delta\rho^{A}+\beta(\Delta\rho^{A})^{2})\frac{\partial}{\partial x}\Delta\rho^{A}\right]\nonumber\\
&-& h_{1}\frac{\partial^{4}}{\partial x^{4}}\Delta\rho^{A} +h_{2}\frac{\partial^{4}}{\partial x^{2}\partial t^{2}}\Delta\rho^{A}\nonumber\\ &+&\nu\frac{\partial^{2}}{\partial x^{2}}\left( \frac{\partial}{\partial t}\Delta\rho^{A}\right) + F_{2}\left( \frac{\Delta\rho^{A}}{\rho^{A}_{0}}, V\right).
\end{eqnarray}

We will now develop the coupling forces between the two models. Note that in the electromechanical model, the input signal is an electrical perturbation that causes a change in membrane potential and propagation of action potential. The action potential exert a force in the lipid membrane which leads to deformation and propagation of density pulse. The tension in the membrane leads to an increase of the transmembranal ion flow and therefore reinforce the AP.
\subsection {Derivation of coupling forces $ F_{1}$ and $ F_{2}$}

The propagation of mechanical waves distends the axon membrane as shown on the figure below.
\begin{figure}[htp]
	\centering
	\includegraphics[width=3.0in,height=2.50in]{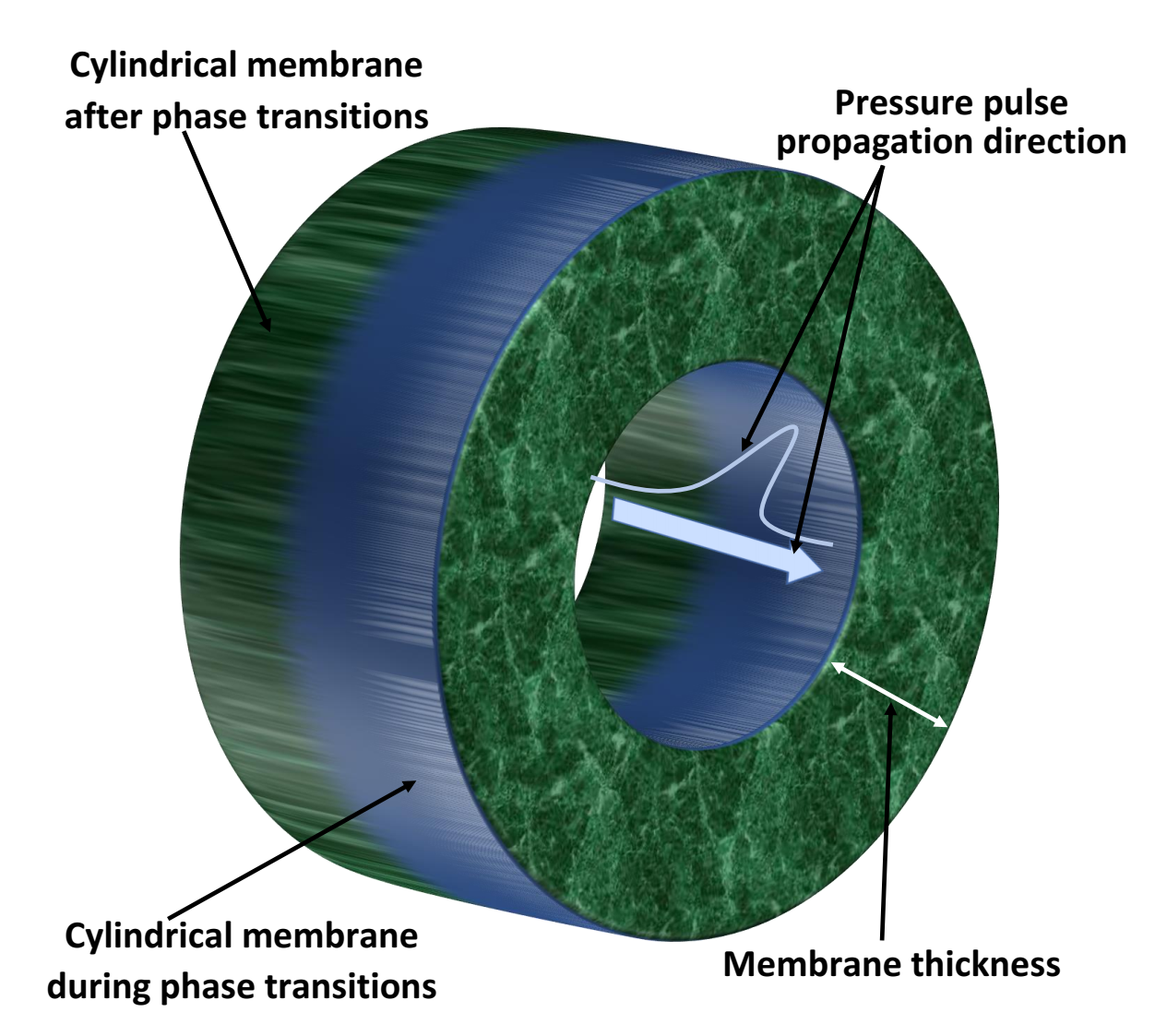}
	\caption{Schematic representation of the radial contraction of the axon membrane due to the axoplasmic pressure pulse. While the lateral density wave propagates along the axon, the longitudinal wave deform the biomembrane. Thus, the electrical properties of the membrane are affected by the application of lateral pressure or tension in a membrane. The charging of the membrane capacitor generates a force on the membrane that leads to an effective reduction of the membrane thickness and thereby to an increase in capacitance with a quadratic voltage dependence.}
	\label{g2}
\end{figure}
During the propagation of the density pulse, the charge $Q$, is stored across the lipid bilayer . The exchange of ions across the membrane generates a potential difference and so the potential energy stored in the cell membrane is given by:
\begin{eqnarray}\label{m1}
E_{c}=\frac{1}{2}Cz^{2} \nonumber 
\end{eqnarray}
where $z$ is the potential difference across the membrane and can also be expressed in terms of charge $Q$ .i.e $z=\frac{Q}{C}$. $C$ is the membrane capacitance which depend on the characteristic of the insulating material between the plates which is the lipids of the plasma membrane
\begin{eqnarray}\label{m2}
C=\frac{\varepsilon A}{\textbf{D}}
\end{eqnarray}
$\varepsilon$ being the relative dielectric constant, $A$ the membrane space area, and \textbf{D } the thickness of the membrane. Considering the properties of viscous compressible fluid,
$\alpha=R\sqrt{\dfrac{\omega \rho^{A}}{\mu}}$  and  $ v_{gr}=\sqrt{\frac{2\omega E\textbf{D}R}{3\mu}}$, we obtain the thickness of the membrane \textbf{D} as a function area density $\rho^{A}$.
\begin{eqnarray}\label{m3}
\textbf{D}=\frac{3Rv_{gr}^{2}\rho^{A}}{2E\alpha^{2}}.
\end{eqnarray} 
From the definition of  $\Delta\rho^{A}=\rho^{A}-\rho_{0}^{A}$, Eq.(\ref{m3}) becomes;
\begin{eqnarray}\label{m4}
\textbf{D}=\frac{3Rv_{gr}^{2}\rho_{0}^{A}}{2E\alpha^{2}}\left( \rho_{0}^{A}+\Delta\rho^{A}\right).
\end{eqnarray}
Given the resting potential $V_{0}$ and $V$ the voltage at a later time t then $z=V_{0}+V$. Assuming that the change in the density of the membrane as the signals propagates is very small i.e $\Delta\rho^{A}<<\rho_{0}^{A}$,  Eq.(\ref{m4}) becomes:
\begin{eqnarray}\label{m5}
E_{c}=\frac{\varepsilon AE\alpha^{2}}{3R v_{gr}^{2}\rho^{A}_{0}}\left[\left(V_{0}+V\right) ^{2}\left( 1- \frac{\Delta\rho^{A}}{\rho^{A}_{0}}\right)\right].
\end{eqnarray}
The initial density of the membrane is given by;
\begin{eqnarray}\label{m6}
\rho_{0}^{A}=\frac{2E\alpha^{2}\textbf{D}_{0}}{3Rv_{gr}^{2}}
\end{eqnarray}
And the initial capacitance of the membrane given by;
\begin{eqnarray}\label{m7}
C_{0}=\frac{\varepsilon A}{\textbf{D}_{0}}
\end{eqnarray}
Substituting Eq.(\ref{m6}) and Eq.(\ref{m7})into Eq.(\ref{m5}), we obtain
\begin{eqnarray}\label{m8}
E_{c}=\frac{C_{0}}{2}\left[(V_{0}+V)^{2}\left( 1-\frac{\Delta\rho^{A}}{\rho^{A}_{0}}\right) \right] \nonumber
\end{eqnarray}

This is the energy stored in the membrane as the signal propagates in the nerve. This energy is a function of voltage  and density describing  the electrical and mechanical phenomena of the  biological system. We can now take a step further to determining the electrical and mechanical coupling forces as:
%
\begin{eqnarray}\label{m91}
F_{1}\left( \frac{\Delta\rho^{A}}{\rho^{A}_{0}}, V\right) =-\frac{\partial E_{c}}{\partial V},\nonumber \\ 
F_{2}\left( \frac{\Delta\rho^{A}}{\rho^{A}_{0}}, V\right) =-\frac{\partial E_{c}}{\partial \Delta\rho^{A}}\nonumber
\end{eqnarray}
wich give:
\begin{eqnarray}\label{m9}
F_{1}(\frac{\Delta\rho^{A}}{\rho^{A}_{0}},V)=-C_{0}[V_{0}-V_{0}\frac{\Delta\rho^{A}}{\rho^{A}_{0}}+V-\frac{\Delta\rho^{A}}{\rho^{A}_{0}}V],\\ \nonumber \\ \nonumber F_{2}(\frac{\Delta\rho^{A}}{\rho^{A}_{0}},V)=\frac{C_{0}}{2\rho^{A}_{0}}(V_{0}^{2}+2V_{0}V+V^{2})
\end{eqnarray}
\subsection{Mechanism of coupling}

Adding  $F_{1}$ into  Eq.(\ref{m110}) and applying $\tau=\frac{t}{\sqrt{LC}}$;  $v=\dfrac{V}{V_{0}}$;  $u=\frac{\Delta \rho^{A}}{\rho_{0^{A}}}$, one obtains a dimensionless form of the FitzHugh Nagumo model equation with a coupling force

\begin{eqnarray}\label{m10}
\dfrac{\partial^{2}v}{\partial\tau^{2}}&=&\gamma\dfrac{\partial v}{\partial\tau}-\dfrac{\gamma}{3}\dfrac{\partial v^{3}}{\partial\tau}+D_{2}\dfrac{\partial^{2} v}{\partial x^{2}} -\dfrac{\beta}{\gamma}\dfrac{\partial v}{\partial\tau}\nonumber\\
&-&\frac{\beta D}{\gamma}\left( 1+v-u-vu\right) \nonumber\\ 
&+& \beta v -\frac{\beta v^{3}}{3}-v+\alpha+D_{0}\dfrac{\partial^{3} v}{\partial x^{2}\tau} \\
&+&D\left( \dfrac{\partial u}{d\tau}-\dfrac{\partial v}{\partial\tau}+v\dfrac{\partial u}{\partial\tau}+u\dfrac{\partial v}{\partial\tau}\right), \nonumber 
\end{eqnarray}
Where $\alpha=\frac{E}{V_{0}}$,\:\:\; $\beta=\frac{R}{\rho}$,\:\:\: $\gamma=\frac{1}{\rho}\sqrt{\dfrac{L}{C}}$, \:\:\: $D=\sqrt{LC}C_{0}$,\:\:\: $D_{2}=CRD_{1}/\sqrt{h_{1}}$,\:\:\: $D_{0}=\sqrt{LC}D_{1}v_{0}c_{0}^{2}/h_{1}$

Also, adding $F_{2}$ into Eq.(\ref{m111}) and applying $u=\frac{\Delta \rho^{A}}{\rho_{0^{A}}}$  ,$x^{'}=\frac{c_{0}x}{\sqrt{h_{1}}}$, one also obtains a dimensionless form of the modified Heimburg Jackson model with a coupling force
\begin{eqnarray}\label{m11}
\frac{\partial^{2}u}{\partial \tau^{2}}&=&\frac{\partial}{\partial x}\left [\left( 1+pu+qu^{2}\right) \frac{\partial u}{\partial x}\right]-H_{1}\frac{\partial^{4}u}{\partial x^{4}}\nonumber\\ &+&H_{1}\frac{\partial^{4}u}{\partial x^{2}\partial \tau^{2}} +\mu\frac{\partial^{3}u}{\partial x^{2}\partial \tau}+\sigma\left( \frac{1}{2}+v+\frac{v^{2}}{2}\right),
\end{eqnarray}
where  
$p=\frac{\rho_{0}^{A}\alpha}{c_{0}}$; $q=\frac{(\rho ^{A}_{0})^{2}\beta}{c_{0}}$; $H_{1}=\frac{1}{h_{1}}$; $H_{2}=h_{2}c_{0}^{2}/h_{1}LC$; $\mu =\frac{\nu}{\sqrt{h_{1}LC}}$, $\sigma=\frac{C_{0}V_{0}^{2}}{\rho_{0}^{A}}$,
$u$ is the localised density pulse, $x$ and $\tau$ are dimensionless variables of space and time respectively where the prime on $x$ has been suppressed for simplicity. $H_{1}$ and $H_{2}$ are the dispersion coefficients, $\sigma$ the coupling coefficient, $\mu$ measures the friction of the nerve axon. Equations (\ref{m10}) and (\ref{m11}) are dimensionless since the dimensional forms of these equations may contain parameters chosen base on the accuracy of the experiment. This coupled equation (Eq (\ref{m10}) and Eq (\ref{m11})) is the electromechanical model of nerves describing ensemble of waves of two physical origins( electrical and mechanical). It is interesting to note that the coupling constants depend on the initial value of the density, the Voltage and the capacitor of the membrane.

It is always a good practice to perform the stability of any system before starting its study and so we are going to proceed  with the stability analysis of our system in the next section.

\section{Stability Analysis}

It should be stress that neurons communicate with each other through the use of electromechanical activities which alters the membrane potential as well as the deformation on the recipient neurons \cite{b31}. To continue propagating this message over a long distance has been proven to be a bit challenging and might require amplification of the signals. Which can lead to instability. Thus, determining whether a given signal is stable is an important tool in neurodynamics. Stability analysis of signals propagating in a nerve axon is also an important issue related to the study of nonlinear dynamical system because it provides an effective way of testing the  robustness of a given signal against small perturbation. In this section, we introduce the coupled-mode system for the propagation of electromechanical signals. Before seeking fixed points of  Eq. (\ref {m10}) and Eq. (\ref{m11}) , we look travelling wave solution of the system by considering change of coordinate $z=x-c_{0}\tau$, where $c_{0}$ is the speed of the signal. Substituting into coupled equations Eq. (\ref{m10}) and Eq. (\ref{m11}),  after reducing to first order differential equations, we obtain a set of seven ordinary differential equations as
\begin{eqnarray}\label{r41}
\dfrac{du}{dz}&=&x\nonumber\\
\dfrac{dx}{dz}&=&y\nonumber\\
\dfrac{dy}{dz}&=&w\nonumber\\
\dfrac{dv}{dz}&=&k \\
\dfrac{dk}{dz}&=&l \nonumber\\
\dfrac{dw}{dz}&=&\frac{1}{\left( H_{1}-c^{2}_{0}H_{2}\right) }\left( (1+pu+qu^{2}-c_{0})y+(p+2uq)x^{2}\right.\nonumber \\
&-&\left.\mu c_{0}w+\sigma\left( \frac{1}{2}+v+\frac{v^{2}}{2}\right)\right)  \nonumber\\
\dfrac{dl}{dz}&=&\frac{1}{D_{0}c_{0}}\left( D_{2}l\right.\nonumber\\
&+&\left.\left( \gamma v^{2}c_{0}-\gamma c_{0}-c^{2}_{0}+\frac{\beta c_{0}}{\gamma}+Dc_{0}-Dc_{0}u\right) k\right.\nonumber\\
&-&\left. Dc_{0}(1+v)x-\frac{\beta D}{\gamma}(1+v-u-uv)\right.\nonumber\\
&+&\left. \left( \beta -\frac{\beta v^{2}}{3}-1\right) v+\alpha\right).\nonumber
\end{eqnarray}

The coupled ordinary differential equations above can be rewritten as a vector field, $\textbf{F(X)}$, of seven components:
\begin{equation}
\frac{d \textbf{X}}{dz}=\textbf{F(X)},\nonumber
\end{equation}
where $\textbf{X}$ represents $(x,y,w,k,l,u,v)$. The fixed points $\textbf{X}_{*}$, are obtain by solving $\textbf{F}(\textbf{X})=0$ simultaneously. The fixed points are given by 
$x_{*}= y_{*}= w_{*}= k_{*}=l_{*}=0$,\quad $v_{*}=-1$,\quad $u_{*}\in\Re$ and so they are an infinite number of fixed points. To determine the stability of a given fixed point, one needs to first construct the Jacobian matrix
\begin{eqnarray}\label{r5}
J(\textbf{X}_{*})=\dfrac{\partial \textbf{F(X)}}{\partial \textbf{X}}.\\ \nonumber \\ \nonumber
\end{eqnarray}
\[\begin{pmatrix}J(\textbf{X}_{*})\end{pmatrix}=\begin{pmatrix}1&0&0&0&0&0&0\\0&1&0&0&0&0&0\\0&0&1&0&0&0&0\\0&0&0&1&0&0&0\\0&0&0&0&1&0&0\\0&m_{1}&m_{2}&0&0&0&0\\0&0&0&m_{3}&m_{4}&0&m_{5}
\end{pmatrix}
\]
with $m_{1}=(1+pu_{*}+qu_{*}^{2}-c_{0})/(H_{1}-c^{2}_{0}H_{2})$,
\:\:\: $m_{2}=\frac{-\mu c_{0}}{(H_{1}-c_{0}^{2}H_{2})}$ ,\:\:\: $m_{3}=(\beta c_{0}/\gamma+Dc_{0}-c^{2}_{0}-Dc_{0}u_{*})/D_{0}c_{0}$,\quad  $m_{4}=\frac{D_{2}}{D_{0}c_{0}}$,\:\:\: $m_{5}=-(\beta D/\gamma-\beta Du_{*}/\gamma+1)/D_{0}c_{0}$.

Next we compute the eigenvalues of ${\textbf{J}(\textbf{X}_{*})}$ by solving the determinant $det\mid{\textbf{J}(\textbf{X}_{*})}-\lambda I\mid=0$

Where $\lambda$ is the eigenvalue and $I$ is the identity matrix. The stability or instability of a fixed point $\textbf{X}_{*}$ is determined by the real parts, $\Re(\lambda)$, of the eigenvalues. If all eigenvalues have real parts less than zero, then the fixed point $\textbf{X}_{*}$ is stable. On the other hand, if at least one of the eigenvalues has a real part greater than zero then the fixed point $\textbf{X}_{*}$ is unstable. If none of the conditions are fulfil, then there is no conclusion.

Proceeding with our analysis as described above, we obtain after some standard calculations the characteristics polynomial
\begin{eqnarray}\label{p2}
a_{0}\lambda^{7}+a_{1}\lambda^{6}+a_{2}\lambda^{5}+a_{3}\lambda^{4}+a_{4}\lambda^{3}+a_{5}\lambda^{2} +a_{6}\lambda=0.\nonumber
\end{eqnarray}

Where, $a_{0}=1$,\:\:\:   $a_{1}=-(m_{5}+5)$,\:\:\: $a_{2}=5m_{5}+10$,\:\:\: $a_{3}=-10m_{5}$,\quad $a_{4}=10m_{5}+5$, $a_{5}=-
(5m_{5}+1)$,\quad$a_{6}=m_{5}$.

We can observe immediately that one of the eigenvalues  is given by $\lambda_{1}=0$. This provides no information about the stability of the system but that the equilibrium points lie in the y-axis. This reduces the characteristic equation to
\begin{eqnarray}\label{p3}
a_{0}\lambda^{6}+a_{1}\lambda^{5}+a_{2}\lambda^{4}+a_{3}\lambda^{3}+a_{4}\lambda^{2}+a_{5}\lambda +a_{6}=0. \nonumber
\end{eqnarray}

The problem of stability analysis using the characteristic equation is difficult if we applied it to a high order system. In this case, it relates to searching the roots of the degree $n$ polynomial equation. With the difficulty of finding the roots of the equation, the stability analysis will be carried out using the Routh-Hurwitz criterion \cite{me, me2}. So this is a means of detecting unstable poles from a characteristic  polynomial without actually calculating  the roots. The characteristic equation is given by Eq. (\ref{p2}). Before we study the Routh-Hurwitz criterion, firstly we will recall the conditions for stable, unstable and marginally stable system.

\begin{enumerate}
	\item If all the roots of the characteristic equation lie on the left half of the complex plane then the system is said to be a stable system.
	\item If all the roots of the system lie on the imaginary axis of the complex plane then the system is said to be marginally stable.
	\item If all the roots of the system lie on the right half of the complex plane then the system is said to be an unstable system.
\end{enumerate}

In the other words the Routh Hurwitz criterion for stability states that any system can be stable if and only if all the roots of the first column have the same sign. If it does not have the same sign or there is a sign change then the number of sign changes in the first column is equal to the number of roots of the characteristic equation in the right half of the s-plane that is equals to the number of roots with positive real parts \cite{me2}. We have to follow  some conditions to make any system stable, or we can say that they are some necessary conditions to make the system  stable. It is worthy to note that this Routh Hurwitz method is valid only for real coefficient \cite{me2, me3}. The Routh Hurwitz table for the characteristic polynomial is as shown below.

\noindent
$\lambda^{6}\quad \quad \quad \quad a_{0}\quad a_{2}\quad a_{4}\quad a_{6}$\\ 
$\lambda^{5}\quad \quad \quad \quad a_{1}\quad a_{3}\quad a_{5}\quad 0$\\ 
$\lambda^{4}\quad \quad \quad \quad b_{1}\quad b_{2}\quad b_{3}$\\ 
$\lambda^{3}\quad \quad \quad \quad c_{1}\quad c_{2}$\\ 
$\lambda^{2}\quad \quad \quad \quad d_{1}\quad d_{2}$\\ 
$\lambda^{1}\quad \quad \quad \quad e_{1}$.

Based on this criterion, we can study the stability of the equilibrium states following some necessary conditions.  This is possible by choosing the parameter $\gamma$(charge density) to satisfy the conditions given in appendix.

Let us determine the range of values of  $\gamma$ (charge density)for which the conditions above are fulfilled by plotting  the conditions $\Delta$ (representing the conditions above) against $\gamma$  as shown in Fig. (\ref{ro}) below.

\begin{figure}[htp]
	\centering
	\includegraphics[width=3in,height=2.0in ,scale=1]{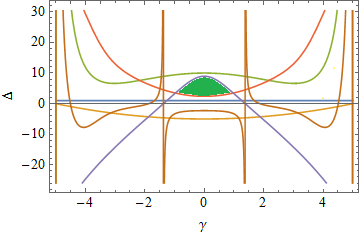}
	\caption{Stability diagram for conditions $\Delta$ against $\gamma$ with parameter values $c_{0}=176.6$, $D_{0}=0.35$, $\beta=0.3$ and step size $h=0.01$. Shaded area(green) is the computed stable region}.
	\label{ro}
\end{figure}

Considering the analytical critical lines related to the Routh-Hurwitz criteria, we identified the range of values of $\gamma$ from $-1$ to $1$  for which the system is stable. Shaded area (green) in Fig. (\ref{ro}) above, is the computed stable region.
Base on these range of $\gamma$ values, let us consider further numerical simulation to investigate the evolution of the fixed points obtained by considering value of $\gamma$ within the stable region and any other value of $\gamma$ out of this stable region.
\section{Numerical Analysis}
In this section, we investigate a number of questions associated with the stability of the fixed points in the stable and unstable regions of Fig. (\ref{ro}) numerically. To this end, it is convenient to consider the first order differential equations as given in Eq. (\ref{r41}).
Performing numerical simulations of this equation through Runge Kutta method of order four, we obtain the following graphs:

\begin{figure}[htp]
	\centering
	\includegraphics[width=5in,height=2in]{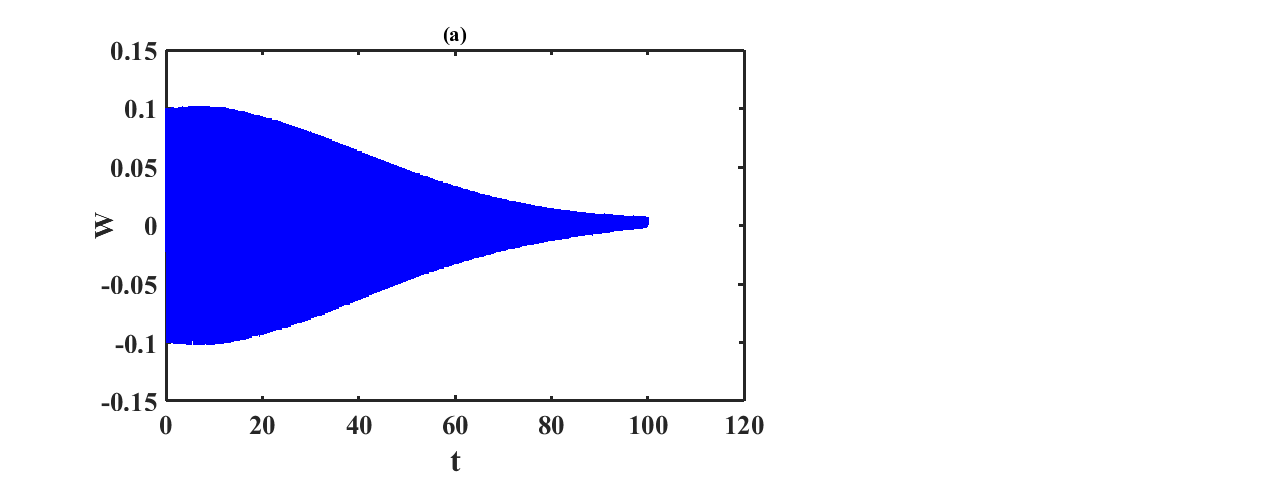}
	\includegraphics[width=5in,height=2in]{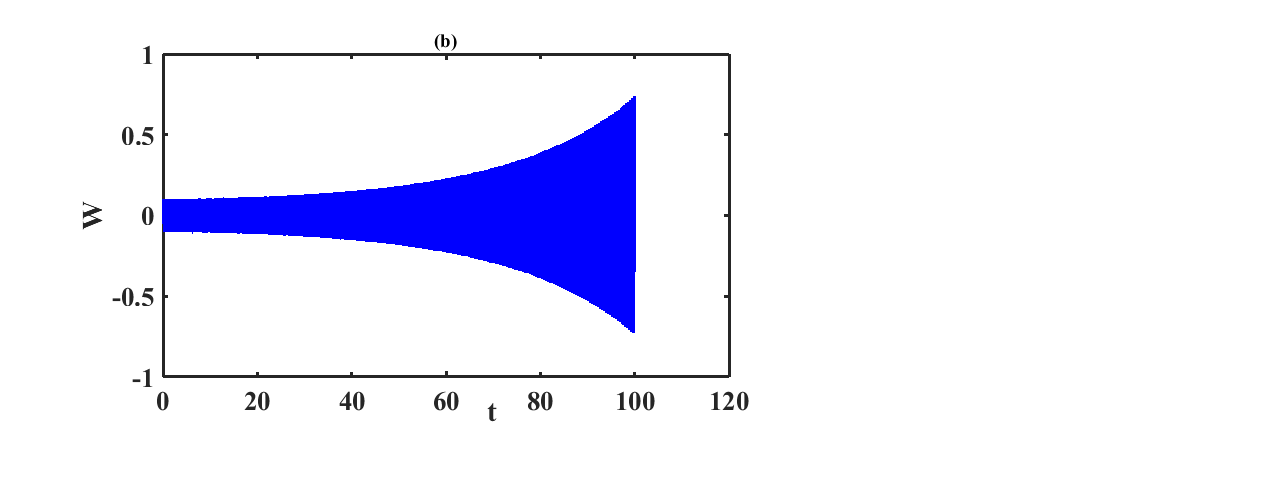}
	\caption{Emergence of stable signals (see Fig. (\ref{nmm}\textbf{a}) for $\gamma=0.05$, unstable signals (see Fig. (\ref{nmm}\textbf{b}), for $\gamma=-2$. other parameter values are: $D=0.095$, $H_1=4$, $H_2=0.5$, $q=79.5$, $p=-16.6$, $\mu=0.045$, $\sigma=0.0086$, $\beta=0.3$, $D_{0}=0.35$, $\alpha=0.45$, $D_{2}=0.095$, $c_{0}=176.6$, and  step size  $h=0.01$.}
	\label{nmm}
\end{figure}
From the time series shown above, its is clear that for the range of parameter values of $\gamma$ in the stable region (shaded in green), the  signal is stable as it is bounded and decays to zero as $t$ grows to infinity as shown in Fig. (\ref{nmm}a) above. On the other side, choosing $\gamma$ from the unstable region (any unshaded region in Fig. (\ref{ro})), we obtained an unstable signal as it grows unbounded  as shown in Fig. (\ref{nmm}b).
\section{Conclusion}
\noindent

With the ongoing argument that the signals propagating in the nerve is governed by electrical variables only or thermodynamic variables only on the other side of the argument, an attempt was made to unify the electrical and soliton model. A model that will help explain the electrical and mechanical activities of nerve signal propagation. We were able to obtain an expression for the energy stored in the biomembrane and corresponding forces as a function of electrical and mechanical variables. From this potential energy,  we were then able to formulate a model that will be able to  describe signals of two physical origins(electrical and mechanical) by unifying the electrical and mechanical models in which the nonlinearity play a decisive role coupled with the simplest idea base on the 'coherent theory' on existing mathematical models of FitzHugh Nagumo  describing the electrical activities of the nerve impulse and the mechanical wave in the surrounding biomembrane govern by the improved Heimburg Jackson model equation. We took a step further to studying the stability of the electromechanical model of nerves through the Routh Hurwitz stability criteria, showing an area for a certain parameter range in which the system is stable on the stability chart. We also carried out numerical simulation making use of the parameter values from the stable and unstable region obtaining stable and unstable signals respectively.

All brain functions are governed by electromechanical signals: it is how memories are stored, habits are made,  pleasure is derived, and how pain is felt. Our model adds to a more complete theory for action potential generation and propagation that explains electrical and  mechanical manifestations of nerve signal propagation. We hope this better theoretical understanding of the phenomena accompanying the action potential will guide experiments seeking to master neuronal information processing. It is worthy to point out that the Routh Hurwitz approach becomes progressively more difficult as the order of the polynomial  increases. But for low-order polynomials, it easily gives stability conditions. We intend to apply perturbation techniques to analyze  in detail this couple model for  better results.

\section*{Appendix}

\begin{widetext}	
\begin{eqnarray}
&&a_{0}=1>0,\\
&&\frac{1}{6\gamma}(D\beta+3\alpha\gamma+4\beta\gamma)-5>0'\\
&&(4\beta \gamma-D\beta+3\alpha \gamma+30\gamma)(-\beta D+3\alpha \gamma+63\gamma) +5(D\beta+3\alpha \gamma+3\gamma+4\beta \gamma)>0\\
&&(\frac{(D\beta+3\alpha \gamma+4\beta \gamma)^{2}}{36\gamma^{2}} -(\frac{10D\beta+3\alpha \gamma+4\beta \gamma-75\gamma}{3\gamma})(\frac{-5(D\beta+3\alpha \gamma+4\beta \gamma)}{3\gamma}))-5\frac{(D\beta+3\alpha \gamma+4\beta \gamma)^{2}}{36\gamma^{2}}\nonumber\\ && +13\frac{(D\beta+3\alpha \gamma+4\beta \gamma+15\gamma)}{3\gamma}-\frac{5}{3}(150\gamma(D\beta+3\alpha \gamma+4\beta \gamma)+(D\beta+3\alpha \gamma+4\beta \gamma)^{2}+180\gamma^{2} >0,
\end{eqnarray}
\begin{eqnarray}
&&(\frac{6\gamma}{D\beta+3\alpha \gamma+4\beta \gamma-30\gamma})((-(\frac{D\beta+3\alpha \gamma+4\beta \gamma}{6\gamma})\\ \nonumber&&+(\frac{-5D\beta-15\alpha \gamma+4\beta \gamma-30\gamma}{6\gamma})(\frac{D\beta+3\alpha \gamma-20\beta \gamma+15\gamma}{3\gamma})))((\frac{5D\beta+15\alpha \gamma+20\beta \gamma-6\gamma}{6\gamma})\\ \nonumber&&-(\frac{D\beta+3\alpha \gamma+4\beta \gamma-30\gamma}{6\gamma})(\frac{-5D\beta-15\alpha \gamma-20\beta \gamma+15\gamma}{3\gamma})+(\frac{5D\beta+15\alpha \gamma+20\beta \gamma}{15\gamma}))\\ \nonumber&&-(\frac{D\beta+3\alpha \gamma+4\beta \gamma-30\gamma}{6\gamma^{2}})(D\beta+3\alpha \gamma+4\beta \gamma)+(\frac{5D\beta+3\alpha \gamma+4\beta \gamma-6\gamma}{6\gamma})>0,
\end{eqnarray}
\begin{eqnarray}
&&\left(\frac{(D\beta+3\alpha \gamma +4\beta \gamma-30\gamma)(D\beta +3\alpha \gamma)}{36\gamma^{2}}+(\frac{5D\beta+3\alpha \gamma+4\beta \gamma-6\gamma}{6\gamma})\right)\\ \nonumber&&\left(\frac{(D\beta+3\alpha \gamma +4\beta \gamma-30\gamma)(D\beta +3\alpha \gamma)}{-36\gamma^{3}}(D\beta+3\alpha \gamma+4\beta \gamma) \right)\\ \nonumber&&-\left(\frac{(5D\beta+15\alpha \gamma +20\beta \gamma-6\gamma)}{6\gamma}+\frac{6\gamma}{(D\beta+3\alpha \gamma +4\beta \gamma-30\gamma)}\right)\\ \nonumber&&\frac{(-5D\beta-15\alpha \gamma -20\beta \gamma+6\gamma)}{6\gamma}\\ \nonumber&&+ \left(\frac{(D\beta+3\alpha \gamma +4\beta \gamma-30\gamma)\left(-5D\beta-15\alpha \gamma -20\beta \gamma-15\gamma\right)}{30\gamma^{2}}\right)\\ \nonumber&&-\left(\frac{(D\beta+3\alpha \gamma +4\beta \gamma-30\gamma)(-5D\beta-15\alpha \gamma -20\beta \gamma-15\gamma)}{30\gamma^{2}}\right)\\ \nonumber&&
-(\frac{6\gamma}{D\beta+3\alpha \gamma +4\beta \gamma-30\gamma})(\frac{-5D\beta-15\alpha \gamma -20\beta \gamma+6\gamma}{6\gamma})\\ \nonumber&&+(\frac{D\beta+3\alpha \gamma +\beta \gamma-30\gamma}{6\gamma})\dfrac{(-5D\beta-15\alpha \gamma -20\beta \gamma+6\gamma)}{6\gamma}\\ \nonumber&& \left(\frac{(5D\beta+15\alpha \gamma +20\beta \gamma-6\gamma)}{6\gamma}-(\frac{(5D\beta+3\alpha \gamma +4\beta \gamma-30\gamma)}{6\gamma}\right).\\ \nonumber&&+\frac{(5D\beta+15\alpha\gamma +20\beta \gamma)}{15\gamma}>0
\end{eqnarray}

\end{widetext}

\end{document}